\documentclass{elsart}

\usepackage{graphicx}
\usepackage{amssymb}

\begin{document}

\begin{frontmatter}



\title{Model-based fit procedure for power-law-like spectra}


\author{Edoardo Milotti}

\address{Dipartimento di Fisica, Universit\`a di Udine and I.N.F.N. -- Sezione di Trieste \\ Via delle Scienze, 208 -- I-33100 Udine, Italy \ead{milotti@ts.infn.it}}

\begin{abstract}
$1/f^\alpha$ noises are ubiquitous and affect many measurements. These noises are both a nuisance and a peculiarity of several physical systems; in dielectrics, glasses and networked liquids it is very common to study this noise to gather useful information. Sometimes it happens that the noise has a power-law shape only in a certain frequency range, and contains other important features, that are however difficult to study because simple fits often fail. Here I propose a model-based fit procedure that performs well on spectra obtained in a molecular dynamics simulation.
\end{abstract}

%MSC:
% 62-07 (STATISTICS: data analysis)
% 62M15 (STATISTICS: inference from stochastic processes: spectral analysis)
% 65D10 (NUMERICAL ANALYSIS: Numerical approximation and computational geometry (primarily 
% algorithms: Smoothing, curve fitting)

%\begin{keyword}

%\PACS 
%\end{keyword}
\end{frontmatter}

\section{Introduction}
\label{intro}

$1/f^\alpha$ noises are very common and affect many measurements; the literature on this subject keeps growing and the apparent ubiquity of these noises has always drawn a great deal of attention. In the experimental practice, they are both a nuisance and a peculiarity of several physical systems; in dielectrics, glasses and networked liquids it is very common to study these noises to gather useful information \cite{sor,isra,CC1,cm}. Sometimes it happens that the noise has a power-law shape only in a certain frequency range which spans several decades, and at the same time contains other important features, that are however difficult to study because simple fits often fail. The main reason of this failure is that the prominent low-frequency peak biases the fit so much that the minute and mostly high-frequency features are neglected. Here I propose a model-based fit procedure that bypasses this problem and that performs well on spectra obtained in a molecular dynamics simulation of water. 

In the rest of this introduction I review the classic superposition argument that relates power-law spectra to the single exponential relaxation processes; in section \ref{seccorr} I analyze the properties of the autocorrelation function of $1/f^\alpha$ spectra; in section \ref{secgallery} I consider the spectral behavior associated to some well-defined distributions of relaxation rates; finally in the last section I show the results of a model-based fit in the case of a molecular dynamics simulation of liquid water, and I summarize my conclusions.

A mathematical mechanism for producing $1/f^\alpha$ noise was proposed long ago by Bernamont \cite{bernamont}, who observed that the superposition of many Lorentzian spectra with a certain distribution of different rates could produce a spectral density with a $1/f$ region. 
The Bernamont superposition argument can be made rigorous with a slight modification of the standard proof of Campbell's theorem \cite{campbell}, and it goes as follows. Take a signal $x(t)$ originated by the linear superposition of many random pulses, i.e., pulses that are random in time and can be described by a memoryless process with a Poisson distribution, have random amplitude $A$ drawn from a distribution with finite variance and probability density  $g_A(A)$, and such that their pulse response function $h(t,\lambda) = \exp(-\lambda t)$ (if $t>0$, otherwise $h(t,\lambda)=0$) is drawn from a distribution with probability density $g_\lambda(\lambda)$. The pulses are received and detected with a rate $n(A,\lambda)$ which in general depends both on the amplitude $A$ and on the decay rate $\lambda$. The pulse arrival process is Poissonian and thus one detects on average $\left[n(A,\lambda) dA d\lambda\right] dt$ pulses in the time interval $(t',t'+dt)$ (and in the amplitude-$\lambda$ range $dA d\lambda$); for the same reason the variance of the number of detected pulses is also equal to $\left[n(A,\lambda) dA d\lambda\right] dt$. This means that the mean square fluctuation of the output signal at time $t$ is given by the integral 
\begin{equation}
\label{msq}
\langle (\Delta x)^2 \rangle = \int_{\lambda_{min}}^{\lambda_{max}} g_\lambda(\lambda) d\lambda \int_{A_{min}}^{A_{max}} g_A(A) dA  \int_{-\infty}^t dt' n(A,\lambda) \left[A h(t-t',\lambda) \right]^2
\end{equation}
If we assume that the rate of occurrence $n$ does not depend on $A$ and $\lambda$, and rearrange the time integration, then the integral (\ref{msq}) simplifies to 
\begin{equation}
\label{msq2}
\langle (\Delta x)^2 \rangle = n \langle A^2 \rangle \int_{\lambda_{min}}^{\lambda_{max}} g_\lambda(\lambda) d\lambda \int_0^\infty dt  \left[ h(t,\lambda) \right]^2
\end{equation}
Now let $H(\omega,\lambda)$ be the Fourier transform of $h(t,\lambda)$, then from the causality constraint on $h(t,\lambda)$ and Parseval's theorem we find that the mean square fluctuation (\ref{msq2}) can be trasformed into 
\begin{eqnarray}
\nonumber
\langle (\Delta x)^2 \rangle &=& \frac{n \langle A^2 \rangle}{2\pi} \int_{\lambda_{min}}^{\lambda_{max}} g_\lambda(\lambda) d\lambda \int_{-\infty}^\infty d\omega  \left| H(\omega,\lambda) \right|^2\\
\label{msq3}
&=& \frac{n \langle A^2 \rangle}{2\pi} \int_{-\infty}^\infty d\omega  \int_{\lambda_{min}}^{\lambda_{max}} g_\lambda(\lambda) d\lambda \left| H(\omega,\lambda) \right|^2
\end{eqnarray}
The right-hand expression in equation (\ref{msq3}) shows that the spectral density is 
\begin{equation}
\label{psd}
S(\omega) = \frac{n \langle A^2 \rangle}{2\pi} \int_{\lambda_{min}}^{\lambda_{max}} g_\lambda(\lambda) d\lambda \left| H(\omega,\lambda) \right|^2
\end{equation}
and since $\left| H(\omega,\lambda) \right|^2=(\omega^2 + \lambda^2)^{-1}$ we obtain eventually 
\begin{equation}
\label{psd2}
S(\omega) = \frac{n \langle A^2 \rangle}{2\pi} \int_{\lambda_{min}}^{\lambda_{max}} \frac{g_\lambda(\lambda)}{\omega^2 + \lambda^2} d\lambda
\end{equation}
If we assume that the decay rates $\lambda$ are uniformly distributed between $\lambda_{min}$ and $\lambda_{max}$ (i.e., $g_\lambda(\lambda) = (\lambda_{max}-\lambda_{min})^{-1}$ ) the spectral density becomes 
\begin{equation}
\label{psdflat}
S(\omega) = \frac{n \langle A^2 \rangle}{2\pi(\lambda_{max}-\lambda_{min})} \frac{1}{\omega} \left(\arctan\frac{\lambda_{max}}{\omega}-\arctan\frac{\lambda_{min}}{\omega}\right)
\end{equation}
so that $S(\omega)$ is approximately constant if $0<\omega\ll\lambda_{min}\ll\lambda_{max}$, and it is approximately equal to 
\begin{equation}
\frac{n \langle A^2 \rangle}{2\pi (\lambda_{max}-\lambda_{min})} \frac{1}{\omega^2}
\end{equation} 
if $\lambda_{min}\ll\lambda_{max}\ll\omega$, while it is approximately equal to 
\begin{equation}
\frac{n \langle A^2 \rangle}{4 (\lambda_{max}-\lambda_{min})} \frac{1}{\omega}
\end{equation}
in the region in between the extreme rates ($\lambda_{min}\ll\omega\ll\lambda_{max}$). 

The spectral density (\ref{psdflat}) has an intermediate region with a $1/f$ behavior, however most observed spectra are not quite $1/f$ but rather $1/f^\alpha$ with $\alpha$ ranging from about 0.5 to nearly 2: how can we obtain such spectra using a superposition as above, i.e., sampling a distribution of  relaxation processes? We could take, e.g., a nonuniform distribution of relaxation processes like $g_\lambda \propto \lambda^{-\beta}$, then in the region $\lambda_{min}\ll\omega\ll\lambda_{max}$ we would find
\begin{eqnarray}
\label{psdbeta}
S(\omega) & \propto & \int_{\lambda_{min}}^{\lambda_{max}} \frac{1}{\omega^2 + \lambda^2} \frac{d\lambda}{\lambda^\beta} =  \frac{1}{\omega^{1+\beta}} \int_{\lambda_{min}/\omega}^{\lambda_{max}/\omega} \frac{1}{1 + (\lambda/\omega)^2} \frac{d(\lambda/\omega)}{(\lambda/\omega)^\beta}\\
& & \approx \frac{1}{\omega^{1+\beta}} \int_0^\infty \frac{1}{1+x^2}\frac{dx}{x^\beta}
\end{eqnarray}

We shall return to these distributions in section \ref{secgallery}.

\section{The rate distribution from the correlation function}
\label{seccorr}
We see that from a given rate distribution we obtain a certain spectral density: can we do the reverse and obtain the rate distribution from a given spectral density? This is not obvious because the spectral density is only a second-order statistics, and does not contain phase information (nor is it possible to preserve it for a noise process). However the answer is yes, the rate distribution can be recovered from the spectral density. 
This can easily be seen from the formal Taylor expansion of the denominator in the integral (\ref{psd2}): 
\begin{equation}
\label{taylor}
S(\omega) = \frac{n \langle A^2 \rangle}{2\pi\omega^2} \int_{\lambda_{min}}^{\lambda_{max}} g_\lambda \sum_{k=0,\infty} \left(-\frac{\lambda}{\omega}\right)^{2k} d\lambda=\frac{n \langle A^2 \rangle}{2\pi\omega^2} \sum_{k=0,\infty}  \left(\frac{-1}{\omega}\right)^{2k} \langle \lambda^{2k}\rangle
\end{equation}
This expansion is only formal inasmuch as it does not converge everywhere, however it shows unequivocally that the shape of $S(\omega)$ depends only on the even moments about the origin of the probability density $g_\lambda$. A probability density function is uniquely determined by the knowledge of {\em all} the moments $\langle \lambda^n\rangle$ (see, e.g., \cite{feller}), and the even moments alone are not enough, but we could still do without the odd moments if the probability density function were an even function.  This is not so, because the decay rates $\lambda$ must be non-negative, and thus the associated probability density function does not have any definite parity. However a probability density function which is non-zero only for positive values of the decay rates can be written in a unique way as the sum of an even and an odd function $g_\lambda(\lambda) = g_\lambda^{(odd)}(\lambda) + g_\lambda^{(even)}(\lambda)$, where $g_\lambda^{(odd)}(\lambda) = g_\lambda^{(even)}(\lambda) = g_\lambda(\lambda)/2$ if $\lambda \ge 0$ and $g_\lambda^{(odd)}(\lambda) = -g_\lambda^{(even)}(\lambda) = -g_\lambda(-\lambda)/2$ if $\lambda < 0$, therefore the odd moments can be computed from the even moments of the distribution, and the even moments alone uniquely identify the rate distribution.

The previous result is only formal and does not yield a practical inversion formula; the actual inversion can be performed in the time domain when we recall that the spectral density $S(\omega)$ is related to the correlation function $R(\tau)$ by the Wiener-Kintchine theorem
\begin{eqnarray}
\label{wk}
R(\tau)&=&\frac{1}{2\pi}\int_{-\infty}^{+\infty} S(\omega) e^{i\omega\tau} d\omega\\
\nonumber
&=&\frac{1}{2\pi} \int_{-\infty}^{+\infty} e^{i \omega \tau} \frac{n\langle A^2\rangle}{2\pi}\int_{\lambda_{min}}^{\lambda_{max}} \frac{g_\lambda(\lambda)}{\omega^2+\lambda^2} d\lambda d\omega\\
\nonumber
&=&\frac{n\langle A^2\rangle}{2\pi}\int_{\lambda_{min}}^{\lambda_{max}} g_\lambda(\lambda) \frac{1}{2\pi}\int_{-\infty}^{+\infty} \frac{e^{i \omega \tau}}{\omega^2+\lambda^2} d\omega d\lambda\\
\label{laplace}
&=&\frac{n\langle A^2\rangle}{2\pi} \int_{\lambda_{min}}^{\lambda_{max}} g_\lambda(\lambda) \frac{e^{-\lambda |\tau|}}{2\lambda} d\lambda
\end{eqnarray}
then we see from equation (\ref{laplace}) that the correlation function is also the Laplace transform of $g_\lambda(\lambda)/(2\lambda)$, and the rate distribution function is uniquely determined by the spectral density and can be retrieved by means of a numerical inverse Laplace transform. In practice, rather than a numerical evaluation of the inverse Laplace transform, one is forced to fit a discrete set of decaying exponentials, and moreover from the correspondence between the Bromwich inversion integral and the inverse Fourier transform, and from the sampling theorem, we see that we must sample the time correlation function, and therefore the noise signal, at a frequency at least twice as high as $\lambda_{max}$ to retrieve $g_\lambda$.  Notice that because of the $\lambda$ in the denominator of the integrand in (\ref{laplace}), the slow relaxations are more heavily weighted in the integral, and the high-frequency parts of the decay rate distribution are much harder to recover than the low-frequency parts; this makes even harder an inversion task which is already known to be very difficult \cite{davismartin}. 

The mixtures of decaying exponentials that characterize many experimental measurements differ significantly only at very short times, while for longer times all the exponentials are equally buried in noise. Disentangling the mixture and finding the relative weights of the different components is possible only if sampling times are very closely spaced at the beginning (and one common strategy is to use logarithmically spaced sampling times (see, e.g. \cite{mx})) and only if one includes some form of prior or assumed knowledge of the distribution of decay rates. There are very few well-established procedures to do this, and the best known are the programs CONTIN and UPEN. CONTIN \cite{sp} uses the following strategies: a) it takes into account {\it absolute prior knowledge}, i.e. whichever exact information that may be available at the beginning, like the non-negativity of decay rates; b) it assumes some {\it statistical prior knowledge} as well, which is essentially the knowledge of the statistics of the measurement noise; c) it uses a {\it principle of parsimony}, which is similar to the principle of maximum entropy, though not as well defined.  UPEN (Uniform PENalty) \cite{upen} assumes instead {\it a priori} that the distribution of decay rates is a continuous function and penalizes distributions which are either discontinuous or have wildly varying curvature. 

In addition to constraints on the shape of the distribution function it is common to use some well-defined standard functions that appear to fit very well many sets of experimental data; the Kohlrausch-Williams-Watts function describes stretched exponentials and works well for relaxations in the time domain and similarly the Havrilijak-Negami (HN) function provides good fits to spectral data. These empirical functions are well-known, and in particular from the HN spectral shape it is possible to compute analytically the distribution of relaxation rates \cite{aac}. However, even though these functions often give satisfactory fits, it would be much better to connect data from experiments or numerical simulations to some well-defined, simple distribution of relaxation rates, just like the spectral density in equation (\ref{psdflat}) can be directly related to a flat distribution of relaxation rates: in the following section I give a list of such spectral shapes. 

\section{A gallery of spectral densities}
\label{secgallery}

\begin{figure}
\begin{center}
\includegraphics[width = 4in]{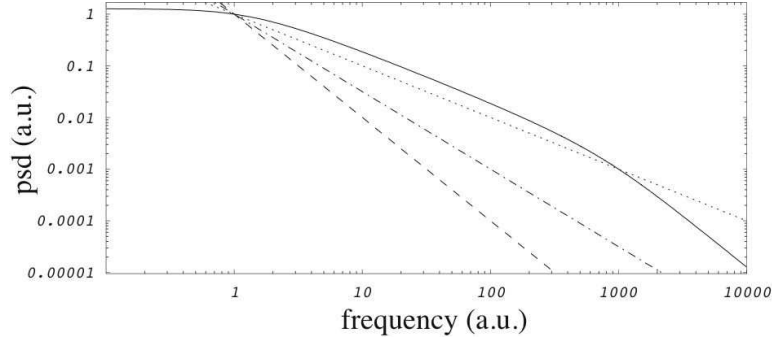}
\caption{Plot of the spectral density (\ref{psdflat}) (solid line); the dotted, dashed-dotted, and dashed lines represent respectively $1/f$, $1/f^{1.5}$, and $1/f^2$ spectra. Both spectral values and frequencies are given in arbitrary units; here $\lambda_{min} = 1 (a.u.)$ and $\lambda_{max} = 1000 (a.u.)$.}
\label{fig1}
\end{center}
\end{figure}

\begin{figure}
\begin{center}
\includegraphics[width = 4in]{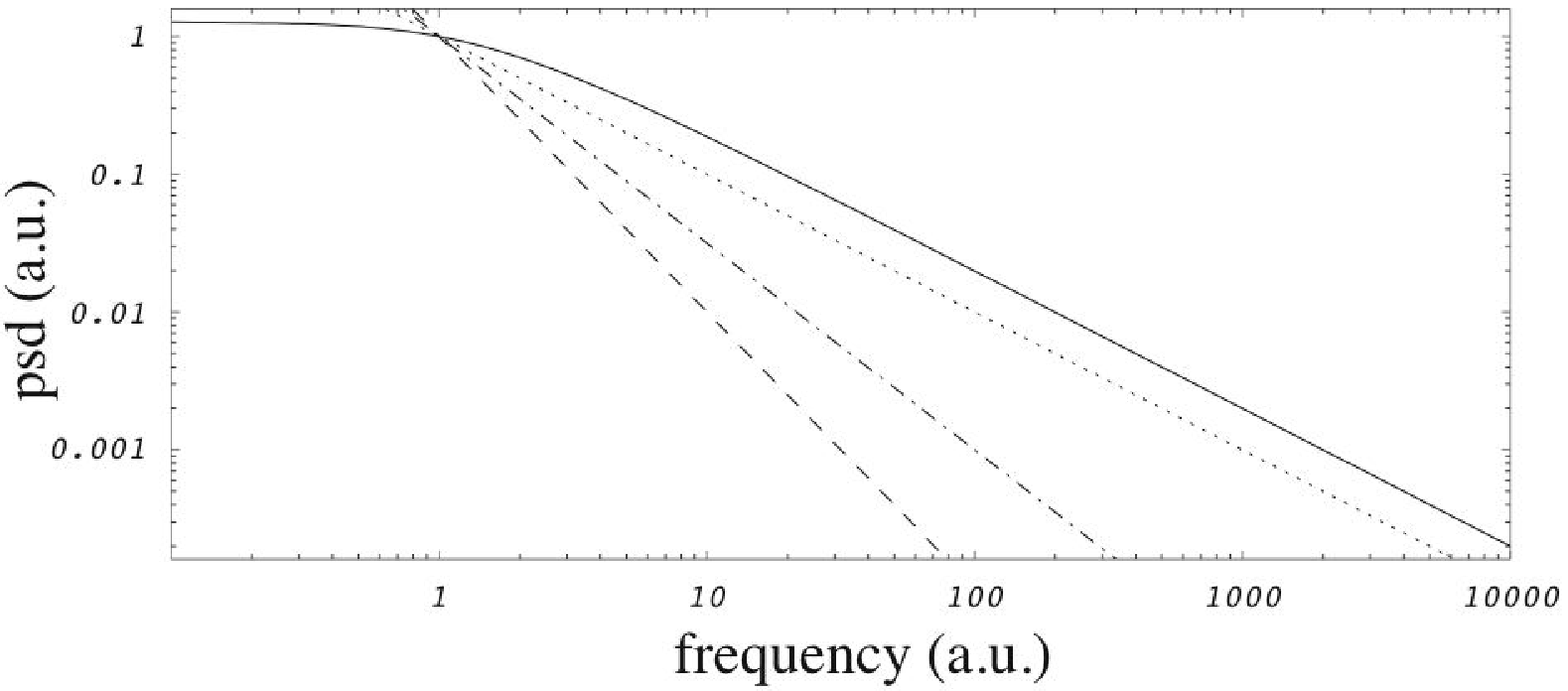}
\caption{Plot of the spectral density (\ref{psdflatlow}) (solid line); the dotted, dashed-dotted, and dashed lines represent respectively $1/f$, $1/f^{1.5}$, and $1/f^2$ spectra. Both spectral values and frequencies are given in arbitrary units; here $\lambda_{min} = 1 (a.u.)$.}
\label{fig2}
\end{center}
\end{figure}

\begin{figure}
\begin{center}
\includegraphics[width = 4in]{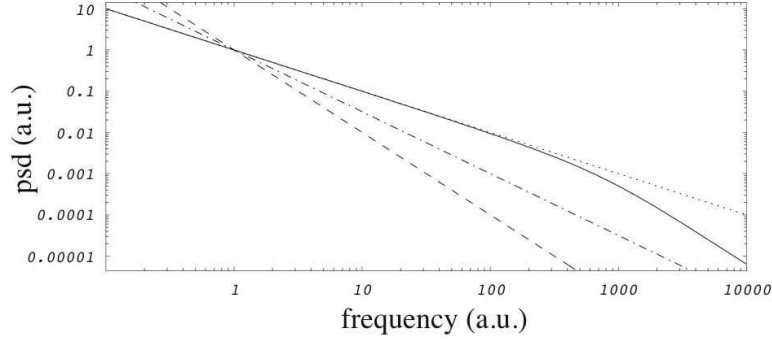}
\caption{Plot of the spectral density (\ref{psdflathigh}) (solid line); the dotted, dashed-dotted, and dashed lines represent respectively $1/f$, $1/f^{1.5}$, and $1/f^2$ spectra. Both spectral values and frequencies are given in arbitrary units; here $\lambda_{max} = 1000 (a.u.)$.}
\label{fig3}
\end{center}
\end{figure}

\begin{figure}
\begin{center}
\includegraphics[width = 4in]{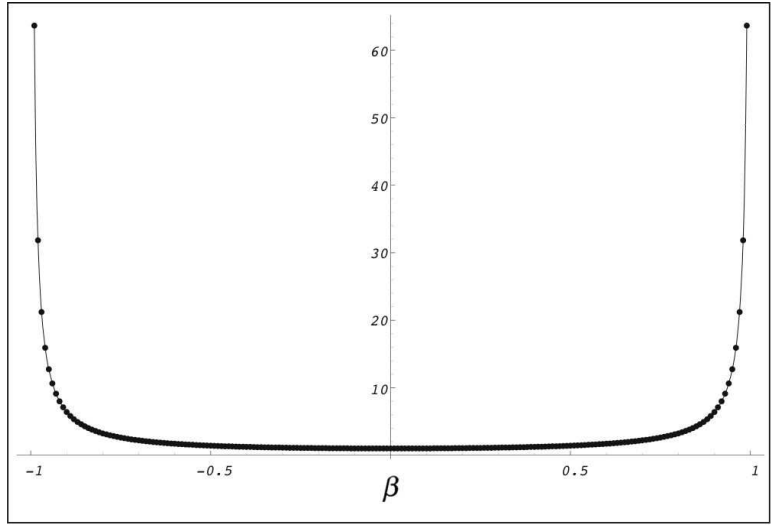}
\caption{Graph of the function $\left[\frac{2\omega^{1+beta}}{\pi} L(\omega,\beta) \right]$ (this product depends on $\beta$ alone); the dots are obtained from numerical estimates of the r.h.s. of equation (\ref{limitL}).}
\label{fig4}
\end{center}
\end{figure}

The spectral density in equation (\ref{psdflat}) produces an intermediate region with a $1/f$ behavior, and includes both a minimum and a maximum relaxation rate: at a frequency lower than the minimum relaxation rate the spectral density whitens and becomes nearly flat, while at a frequency higher than the maximum relaxation rate the spectral density bends downward and assumes a $1/f^2$ behavior, and for fitting purposes we define the standard spectral density
\begin{equation}
\label{psdflatflat}
S_{\rm flat}(\omega; \lambda_{min}, \lambda_{max}) = \frac{1}{\omega} \left(\arctan\frac{\lambda_{max}}{\omega}-\arctan\frac{\lambda_{min}}{\omega}\right)
\end{equation}
However either the minimum or the maximum relaxation rate (or both) may be out of the experimental or numerical simulation range: in these cases the bends at low- and high-frequency become invisible, and a fit with the spectral density (\ref{psdflatflat}) is unstable (at least one of the range parameters is invisible and the chi-square hypersurface flattens out in that direction, adversely influencing the fit). This can be corrected using the modified spectral density
\begin{equation}
\label{psdflatlow}
S_{\rm flat,A}(\omega; \lambda_{min}) = \frac{1}{\omega} \left[\frac{\pi}{2} - \arctan\left(\frac{\lambda_{min}}{\omega}\right)\right]
\end{equation}
when the maximum observable frequency is smaller than the maximum relaxation rate (and the minimum relaxation rate is in the observable range). We should use instead the spectral density 
\begin{equation}
\label{psdflathigh}
S_{\rm flat,B}(\omega; \lambda_{max}) = \frac{1}{\omega}  \arctan\left(\frac{\lambda_{max}}{\omega}\right)
\end{equation}
when the minimum observable frequency is higher than the minimum relaxation rate (and the maximum observable rate is in the observable range), and finally the spectral density 
\begin{equation}
\label{psd1overf}
S_{\rm 1overf}(\omega) = \frac{1}{\omega}
\end{equation}
when both the minimum and the maximum relaxation rates are out of range; the spectral densities (\ref{psdflat}), (\ref{psdflatlow}), and (\ref{psdflathigh}) are shown in figures \ref{fig1} to \ref{fig4}. Using (\ref{psdflatlow}), (\ref{psdflathigh}) or (\ref{psd1overf}) improves the fit stability but means that the final description of the relaxation rate distribution is incomplete.

We have already given a simple argument that shows that a nonuniform distribution of relaxation processes like $g_\lambda \propto \lambda^{-\beta}$ between the maximum and minimum relaxation rates $\lambda_{min}$, $\lambda_{max}$, produces a spectral density with an intermediate $1/f^{1+\beta}$ region: an exact integration yields the spectral density 
\begin{eqnarray}
\label{psdbetaexact}
\nonumber
S_{\rm pl}(\omega; \lambda_{min}, \lambda_{max}, \beta) &=& \frac{1}{(1-\beta)\omega^2}\left[  \lambda_{max}^{1-\beta} F\left(\frac{1-\beta}{2},1;\frac{1-\beta}{2};\frac{-\lambda_{max}^2}{\omega^2}\right) \right. \\ && \left.- \lambda_{min}^{1-\beta} F\left(\frac{1-\beta}{2},1;\frac{1-\beta}{2};\frac{-\lambda_{min}^2}{\omega^2}\right)  \right]
\end{eqnarray}
where $F(a,b;c;z) = \sum_{k=0}^{\infty} \frac{(a)_k (b)_k}{(c)_k} \frac{x^k}{k!}$ is the hypergeometric function and $\beta \in (-1,1)$. Just as in the $1/f$ case either the minimum or the maximum relaxation rate (or both) may be out of the experimental or numerical simulation range and a fit with the spectral density (\ref{psdbetaexact}) becomes unstable, and this can be corrected with the modified spectral densities
\begin{equation}
\label{psdbetalow}
S_{\rm pl,A}(\omega; \lambda_{min}, \beta) = L(\omega,\beta) - \frac{1}{(1-\beta)\omega^2}\left[  \lambda_{min}^{1-\beta} F\left(\frac{1-\beta}{2},1;\frac{1-\beta}{2};\frac{-\lambda_{min}^2}{\omega^2}\right)  \right]
\end{equation}
when the maximum observable frequency is smaller than the maximum relaxation rate (and the minimum relaxation rate is in the observable range) and where the function
\begin{equation}
\label{limitL}
L(\omega,\beta) = \lim\limits_{\lambda_{max} \to \infty}   \frac{\lambda_{max}^{1-\beta}}{(1-\beta)\omega^2} F\left(\frac{1-\beta}{2},1;\frac{1-\beta}{2};\frac{-\lambda_{max}^2}{\omega^2}\right)
\end{equation}
is shown in figure \ref{fig4} and is well approximated by the rational function 
\begin{equation}
\frac{\pi}{2\omega^{1+beta}} \frac{1}{\left( 1+ c_2 \beta^2 + c_4 \beta^4 + c_6 \beta^6 + c_8 \beta^8 + c_{10} \beta^{10} \right)}
\end{equation}
with 
\begin{enumerate}
\item $c_2 \approx -1.2337$
\item $c_4 \approx 0.253669$
\item $c_6 \approx -0.0208621$
\item $c_8 \approx 0.000917057$
\item $c_{10} \approx -0.0000235759$
\end{enumerate}

\begin{figure}
\begin{center}
\includegraphics[width = 4in]{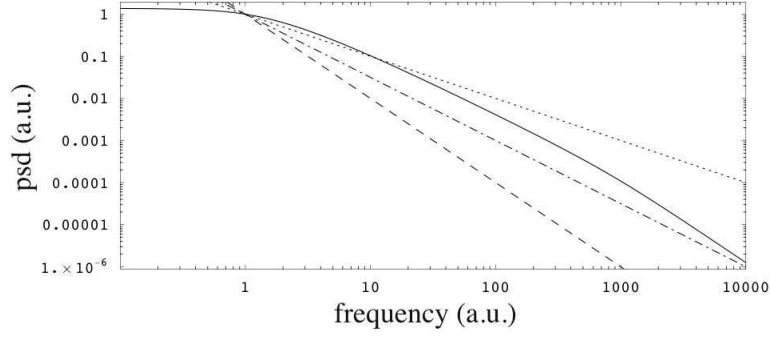}
\caption{Plot of the spectral density (\ref{psdbetaexact}) (solid line); the dotted, dashed-dotted, and dashed lines represent respectively $1/f$, $1/f^{1.5}$, and $1/f^2$ spectra. Both spectral values and frequencies are given in arbitrary units; here $\lambda_{min} = 1 (a.u.)$, $\lambda_{max} = 1000 (a.u.)$, and $\beta = 0.5$.}
\label{fig5}
\end{center}
\end{figure}

\begin{figure}
\begin{center}
\includegraphics[width = 4in]{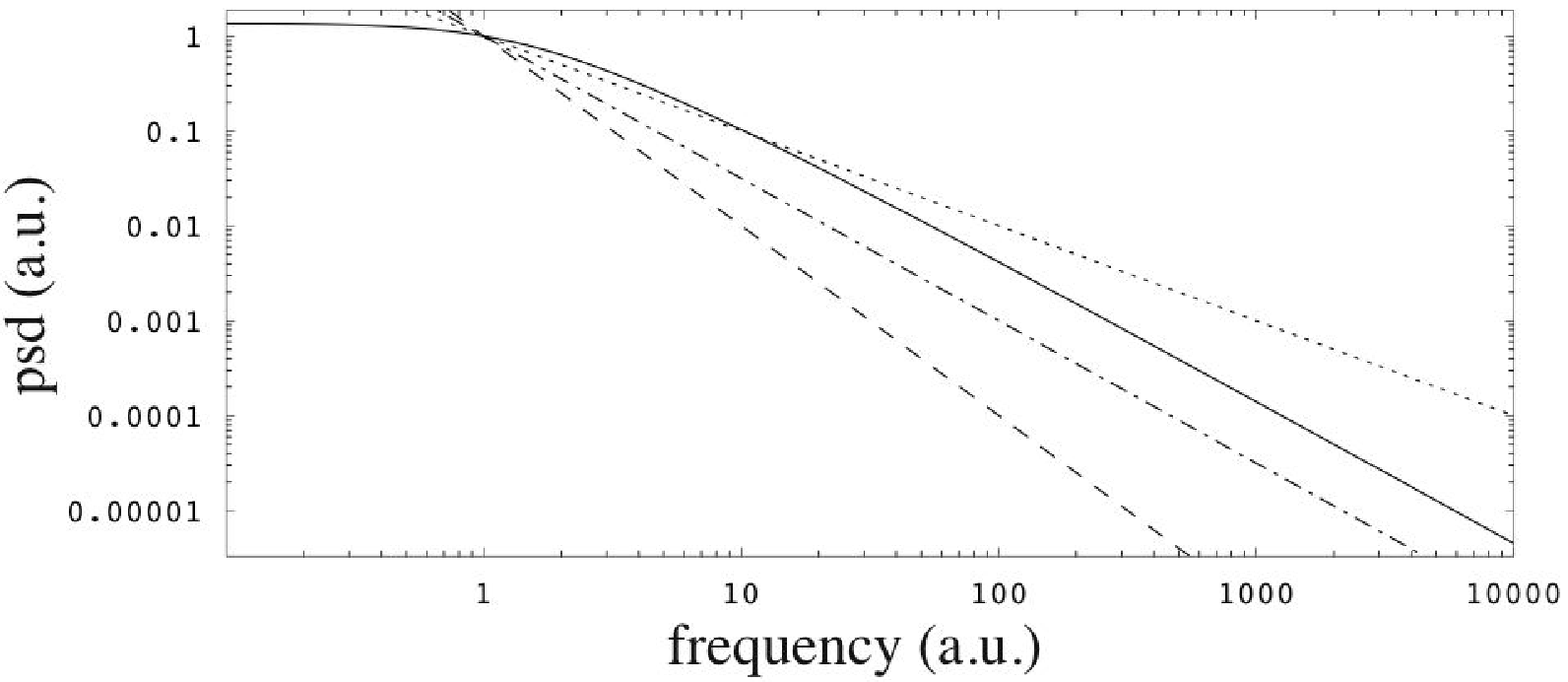}
\caption{Plot of the spectral density (\ref{psdbetalow}) (solid line); the dotted, dashed-dotted, and dashed lines represent respectively $1/f$, $1/f^{1.5}$, and $1/f^2$ spectra. Both spectral values and frequencies are given in arbitrary units; here $\lambda_{min} = 1 (a.u.)$, and $\beta = 0.5$.}
\label{fig6}
\end{center}
\end{figure}

\begin{figure}
\begin{center}
\includegraphics[width = 4in]{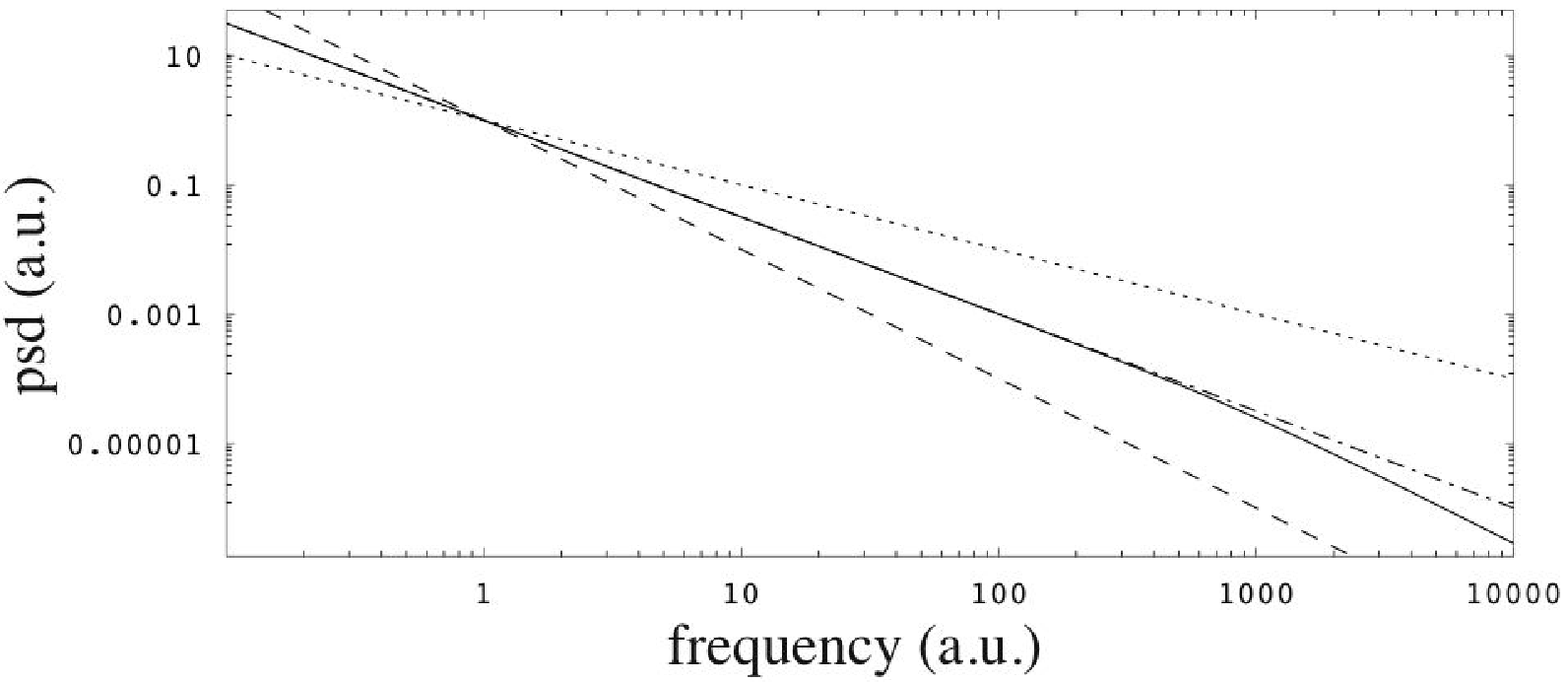}
\caption{Plot of the spectral density (\ref{psdbetahigh}) (solid line); the dotted, dashed-dotted, and dashed lines represent respectively $1/f$, $1/f^{1.5}$, and $1/f^2$ spectra. Both spectral values and frequencies are given in arbitrary units; here $\lambda_{max} = 1000 (a.u.)$, and $\beta = 0.5$.}
\label{fig7}
\end{center}
\end{figure}

The spectral density
\begin{equation}
\label{psdbetahigh}
S_{\rm pl,B}(\omega; \lambda_{max}, \beta) =  \frac{\lambda_{max}^{1-\beta} }{(1-\beta)\omega^2} F\left(\frac{1-\beta}{2},1;\frac{1-\beta}{2};\frac{-\lambda_{max}^2}{\omega^2}\right)
\end{equation}
works when the minimum observable frequency is higher than the minimum relaxation rate (and the maximum observable rate is in the observable range), and finally the spectral density 
\begin{equation}
\label{psd1overfbeta}
S_{\rm 1overf}(\omega; \beta) \propto \frac{1}{\omega^{1+\beta}}
\end{equation}

when both the minimum and the maximum relaxation rates are out of range (here I extend the notation of definition (\ref{psd1overf}) ); the spectral densities (\ref{psdbetaexact}), (\ref{psdbetalow}), and (\ref{psdbetahigh}) are shown in figures \ref{fig5} to \ref{fig7}.

In addition to these distributions, it is possible to consider other shapes like $g_\lambda(\lambda) \propto a + b \lambda$ so that the resulting spectral density from equation (\ref{psd2}) is the sum of a spectral density like the one in equation (\ref{psdflat}) plus a term proportional to 
\begin{equation}
\int_{\lambda_{min}}^{\lambda_{max}} \frac{\lambda}{\lambda^2+\omega^2} \propto \ln\frac{\lambda^2_{max}+\omega^2}{\lambda^2_{min}+\omega^2};
\end{equation}
but I shall not consider them here, since these shapes seem to be far less common than the cases discussed above.

\begin{figure}
\begin{center}
\includegraphics[width = 4in]{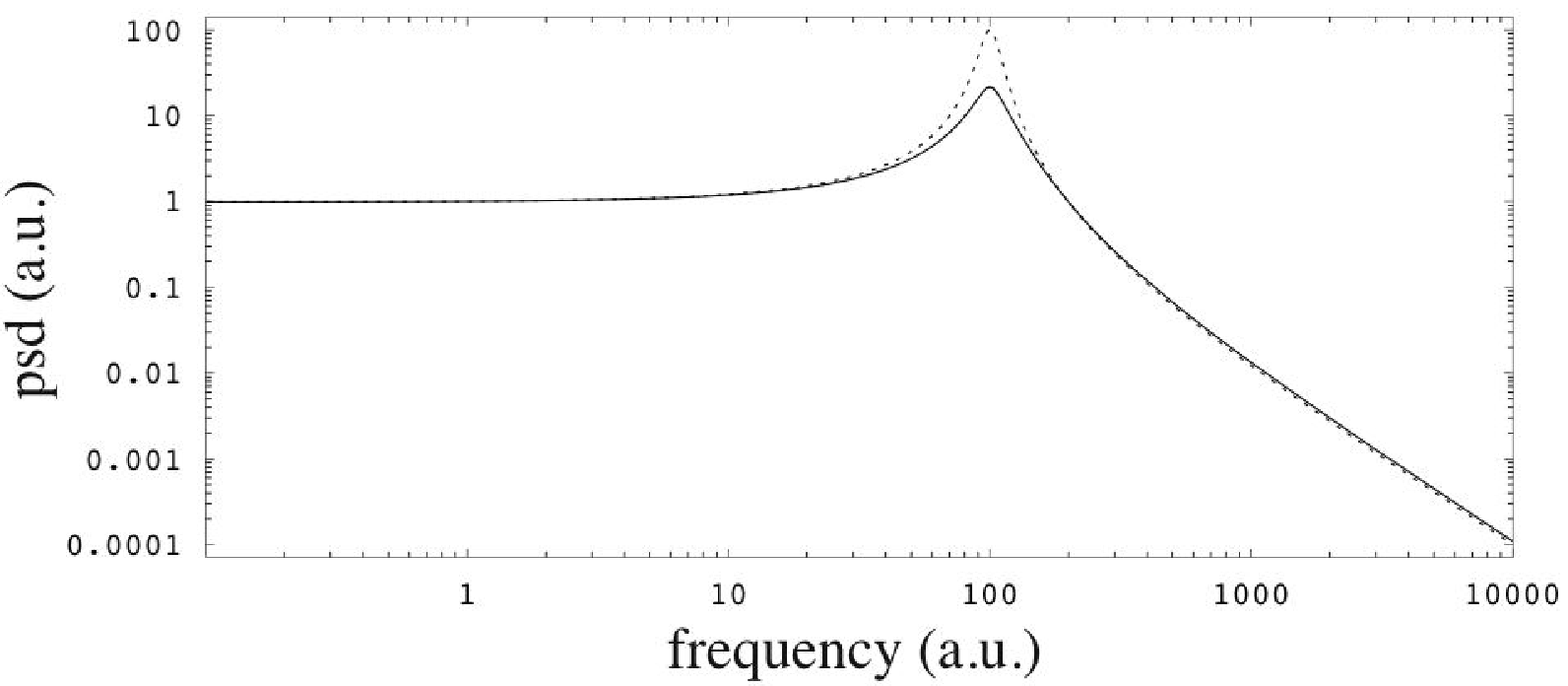}
\caption{Plot of the spectral density (\ref{resof}) (solid line); the dotted, line represents a simple resonance. Both spectral values and frequencies are given in arbitrary units; here $\omega_0 = 100 (a.u.)$,  $\lambda_{min} = 10 (a.u.)$ and $\lambda_{max} = 50 (a.u.)$, while the simple resonance has $\omega_0 = 100 (a.u.)$ and $\lambda = 10 (a.u.)$.}
\label{fig8}
\end{center}
\end{figure}

\begin{figure}
\begin{center}
\includegraphics[width = 4in]{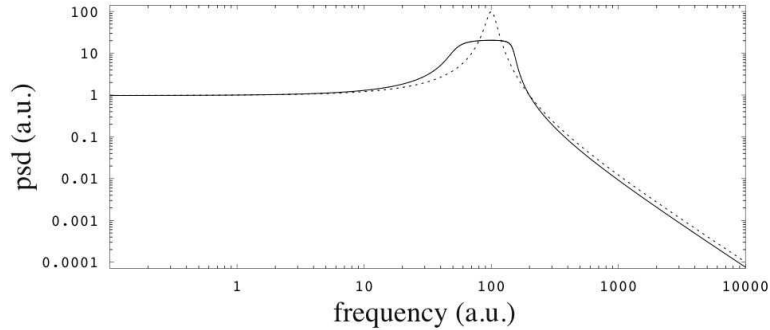}
\caption{Plot of the spectral density (\ref{reslf}) (solid line); the dotted, line represents a simple resonance. Both spectral values and frequencies are given in arbitrary units; here $\lambda = 10 (a.u.)$,  $\omega_{min} = 50 (a.u.)$ and $\omega_{max} = 150 (a.u.)$, while the simple resonance has $\omega_0 = 100 (a.u.)$ and $\lambda = 10 (a.u.)$.}
\label{fig9}
\end{center}
\end{figure}

The integral (\ref{psd2}) is a sum of functions that decrease for positive, increasing $\omega$ and therefore cannot be an increasing function and therefore no distribution of relaxation rates can possibly describe bumps and other small local features such as those that are observed in the spectral densities of glassy systems. These features can be described by resonances or by groups of close resonances; the simplest choices are a) fixed resonance frequency and flat distribution of relaxation rates; b) fixed relaxation rate and flat distribution of resonance frequencies. In the case of a flat distribution of relaxation rates between the maximum and minimum rates $\lambda_{min}$, $\lambda_{max}$ we find
\begin{eqnarray}
\nonumber
S_{\rm fr}(\omega; \lambda_{min}, \lambda_{max}, \omega_0) &=& \int_{\lambda_{min}}^{\lambda_{max}} \frac{d\lambda}{\lambda^2 + (\omega-\omega_0)^2} \\ 
\label{resof}
&=& \frac{1}{\omega-\omega_0}\left[ \arctan\frac{\lambda_{max}}{\omega-\omega_0}-\arctan\frac{\lambda_{min}}{\omega-\omega_0} \right]
\end{eqnarray}
and similarly in the case of a flat distribution of resonance frequencies between the maximum and minimum frequencies $\omega_{min}$, $\omega_{max}$ we find
the spectral densities (\ref{resof}) and (\ref{reslf}) are shown in figures \ref{fig8} and \ref{fig9}.
\begin{eqnarray}
\nonumber
S_{\rm fw}(\omega; \omega_{min}, \omega_{max}, \lambda) &=& \int_{\omega_{min}}^{\omega_{max}} \frac{d\omega_0}{\lambda^2 + (\omega-\omega_0)^2} \\ 
\label{reslf}
&=& \frac{1}{\lambda}\left[ \arctan\frac{\omega - \omega_{min}}{\lambda}-\arctan\frac{\omega - \omega_{max}}{\lambda} \right];
\end{eqnarray}

\section{Model-based fit of a simulated spectral density}
\label{secresults}

When fitting spectra it is important to include the variance of spectral data: if $S_k$ is the spectral estimate at the {\it k}-th frequency, and if the time-domain data are affected by Gaussian white noise, then the spectral estimate of the white noise background has standard deviation $S_k$ \cite{specvar}; this estimate of the standard deviation is usually assumed for simplicity even when there are deterministic components or the noise is not white. Moreover if the final spectral density is the average of $M$ uncorrelated spectra, then the estimate of the standard deviation at the {\it k}-th frequency is  $S_k/\sqrt{M}$. I wish to stress that this treatment of the spectral variance is only approximate in the case of colored noises, but it is assumed nonetheless, because of the complexity of a calculation that includes the correlation between different samples in the time domain (see, e.g. \cite{tm}).

\begin{figure}
\begin{center}
\includegraphics[width = 5in]{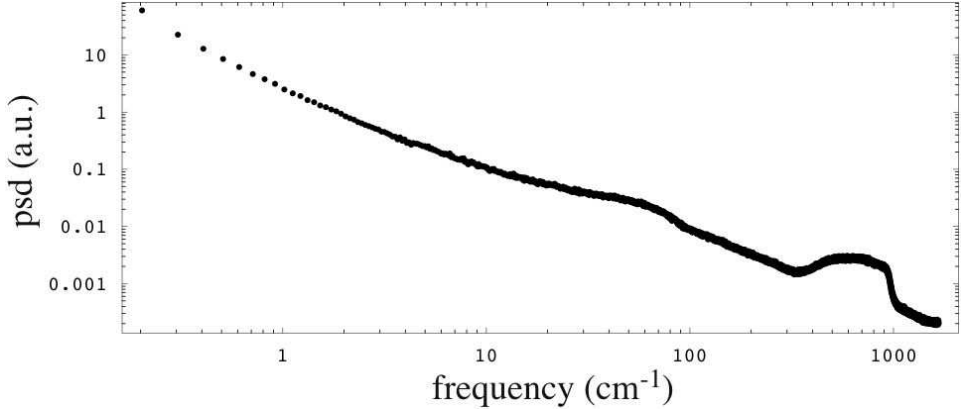}
\caption{Spectral data from \cite{chmu} (230 K data in figure 1a): mean square fluctuation of potential energy vs. frequency. The overall shape is close to a $1/f^\alpha$ spectrum, but notice the low-frequency steepening of the spectrum and the pair of bumps: the low-frequency steepening can be associated to a strong single relaxation, while the bumps should correspond to two resonance distributions like those in equation (\ref{reslf}).}
\label{fig10}
\end{center}
\end{figure}

\begin{figure}
\begin{center}
\includegraphics[width = 5in]{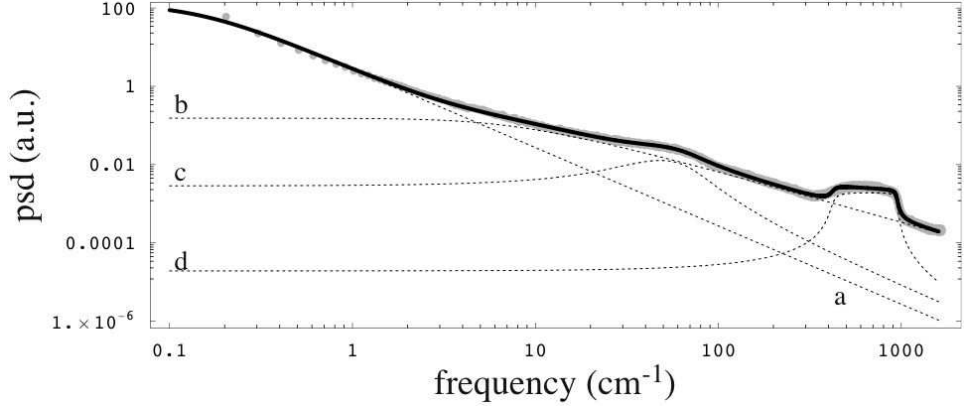}
\caption{Fit to the spectral data from \cite{chmu} shown in figure \ref{fig10} (thick black curve). The data are shown in light gray in the background; the dotted curves {\em a}, {\em b}, {\em c}, and {\em d} represent respectively the first, second, third and fourth term of the model (\ref{model}).}
\label{fig11}
\end{center}
\end{figure}

I have tested the simple model-derived spectral densities described in section \ref{secgallery} on data kindly provided by C. Chakravarty and A. Mudi \cite{chmupri}: the original spectral data are shown in figure \ref{fig10} and correspond to the 230 K curve in figure 1a of reference \cite{chmu} (see also \cite{chmu1,chmu2,chmu3} for full simulation details).
At very low frequency the spectrum is rather steep: a simple fit of the low-frequency data shows a $1/f^2$ behavior, and thus we can surmise that this is just the high-frequency tail of a simple relaxation with a very low relaxation constant (this accounts for 2 fit parameters: amplitude and relaxation rate). At higher frequency the slope is smaller and Mudi and Chakravarty estimate a spectral index slightly higher than 1 \cite{chmu}: since there is no hint of a downward bend, I exclude the full spectral shape (\ref{psdbetaexact}) and also the reduced form (\ref{psdbetahigh}), and I choose (\ref{psdbetalow}) instead, i.e. I include the possibility of a low-frequency flattening, made invisible by the high-frequency tail of the simple relaxation (this adds three more parameters to the fit: an amplitude, a minimum relaxation rate, and a spectral index $\beta$). The high-frequency bump resembles rather closely the shape in figure \ref{fig9}, and thus it is reasonable to assume that both the low-frequency and the high-frequency bumps correspond to flat superpositions of resonances like in equation (\ref{reslf}) (each bump accounts for 4 more parameters: an amplitude, a relaxation rate, a minimum and a maximum resonance frequency, but the relaxation rate is taken to be the same in both bumps). The resulting 12 parameter model is:
\begin{eqnarray}
\nonumber
S(\omega) &=& \frac{a_1^2}{\omega^2 + \lambda_1^2} + a_2^2 S_{\rm pl,A}(\omega; \lambda_{min,2}, \beta)  \\ 
\label{model}
&& + a_3^2 S_{\rm fw}(\omega; \omega_{min,3}, \omega_{max,3}, \lambda_{34}) + a_4^2 S_{\rm fw}(\omega; \omega_{min,4}, \omega_{max,4}, \lambda_{34})
\end{eqnarray}
Notice that the assumptions on the relaxation rate distributions help keep the number of fit parameters rather low. If we tried to fit with a superposition of $N$ simple relaxations we would have $2N$ parameters (one amplitude plus one relaxation rate for each relaxation component): with 12 parameters we could fit only 6 simple relaxation components, therefore the assumed shapes (that correspond to given distributions of relaxation rates and resonance frequencies) allow for a much more economical fit procedure. In this case the spectral data are averages of $M=448$ spectra; table \ref{tab1} lists the fit parameters to the data \cite{chmupri} obtained with a standard Levenberg-Marquardt chi-square minimization procedure, and figure \ref{fig11} compares the fit with the data (the $a$ amplitude values in the table are in the spectral amplitude units of fig. \ref{fig10}, the $\lambda$'s and the $\omega$'s are in cm$^{-1}$, and $\beta$ is dimensionless).

\begin{table*}[htb]
\caption{Fit parameters for the model in equation (\ref{model}) to the data from \cite{chmu}}
\label{tab1}
\begin{tabular}{ll}
\hline 
$a_1$ & $11.435 \pm 0.880$ \\
$\lambda_1$ & $0.144 \pm 0.012$ \\
$a_2$ & $1.351 \pm 0.0059$ \\
$\lambda_{min,2}$ & $5.226 \pm 0.099$ \\
$\beta$ & $0.327 \pm 0.013$ \\
$a_3$ & $0.102 \pm 0.002$ \\
$\omega_{min,3}$ & $32.1 \pm 1.0$ \\
$\omega_{max,3}$ & $64.4 \pm 0.5$ \\
$a_4$ & $0.02534 \pm 0.00002$ \\
$\omega_{min,4}$ & $421.0 \pm 0.2$ \\
$\omega_{max,4}$  & $947.2 \pm 0.1$ \\
$\lambda_{34}$ & $22.77 \pm 0.11$ \\
\hline \\
\end{tabular}
\end{table*}

\begin{figure}
\begin{center}
\includegraphics[width = 5in]{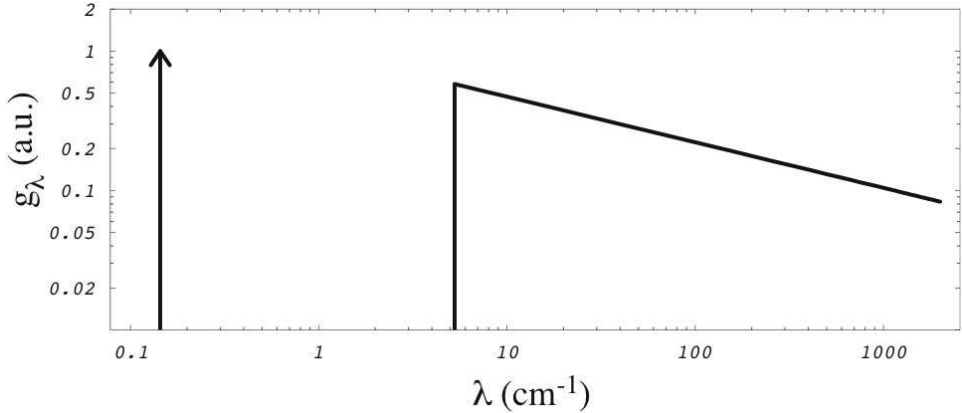}
\caption{Projection of the two-parameter distribution $g(\lambda, \omega_0)$ that describes the model  (\ref{model}) on the $\lambda$ axis. The single relaxation corresponds to a delta-function (arrow on the left).}
\label{fig12}
\end{center}
\end{figure}

The model (\ref{model}) is a function of both relaxation rate and resonance frequency and should thus be described by a two-parameter distribution $g(\lambda, \omega_0)$ rather than $g_\lambda(\lambda)$, however if we concentrate on the projection on the $\lambda$ axis, then we can consider only the first two terms: the (reduced) $\lambda$ distribution is shown in figure \ref{fig12}, and is the sum of a delta-function plus an (unbounded) continuous distribution. Notice that such a distribution is quite challenging for other fit methods, like those implemented by CONTIN and UPEN.

\section{Conclusion}
In this paper I have described a model-based fit of power-law-like spectral densities. Like other similar methods, it embodies {\it a priori} information on the shape of the distribution, but unlike the other methods, the shape is physically motivated, and the fits can be efficiently performed with a reduced number of parameters.

\ack

I wish to thank Giorgio Careri, Giuseppe Consolini, and Charusita Chakravarty for useful discussions. I also wish to thank Charusita Chakravarty and Anirban Mudi for allowing me to use the spectral data from their extensive molecular dynamics simulations of water.

\end{document}